# A Process-driven View on Summative Evaluation of Visual Analytics Solutions

Mosab Khayat        Arif Ghafoor

**Purdue University**

**Abstract**

Many evaluation methods have been applied to assess the usefulness of visual analytics solutions. These methods are branching from a variety of origins with different assumptions, and goals. We provide a high-level overview of the process employed in each method using the generic evaluation model "GEM" that generalizes the process of usefulness evaluation. The model treats evaluation methods as processes that generate evidence of usefulness as output. Our model serves three purposes: It educate new VA practitioners about the heterogeneous evaluation practices in the field, it highlights potential risks in the process of evaluation which reduces their validity and It provide a guideline to elect suitable evaluation method.

**Introduction**

Visual analytics solutions emerged in the past decade and tackled many problems in a variety of domains. The beauty of the idea of combining the power of human and machine in data analysis problems creates fertile ground for new solutions to grow. However, the rise of these hybrid solu-tions complicates the process of evaluation. Unlike automated algorithmic solutions, visual analyt-ics solutions' behavior depends on the user who operates them. This creates a new dimension of variability in the performance of the solutions that need to be accounted for in summative evalua-tion. The existence of a human in the loop, on the other hand, allows researchers to reuse evalua-tion methods from domains, such as anthropology, to extract information with the help of the user. Such evaluation methods allow developers to assess their solutions even when formal summative evaluation is not feasible. The challenge in these methods, however, appears in gathering and ana-lyzing qualitative data to build evidence that are valid.

Many methods have been applied to evaluate visual analytics solutions. Each of these methods was traditionally developed to answer different questions with different evaluation intentions including formative, summative and exploratory 1. Nevertheless, many of these methods have been exten-sively applied in summative evaluation contexts despite the original intention of the methods. In this paper, we abstract the discussion about summative evaluation to reason about the selection of evaluation method for a given evaluation task. We argue that the selection depends on the feasibility of using a method and the quality of the evidence it generates. To diagnose these factors in evalua-tion methods that differ in the core principles of evaluation, a unified system is needed. That system should clarify similarities and differences between the process of evaluation employed in different methods.

To enrich the discussion about this idea of unified system, we propose the Generic Evaluation Model "GEM". This model is built with an illustrative goal to improve the understanding of the feasibility of different evaluation methods and the quality of the evidence they generate. To build the model, we capture common evaluation methods that have been applied to evaluate the useful-ness of visual analytics solutions. Then, we develop the GEM by examining the process of evalua-tion in each method and categorizing the methods based on the similarity of their processes. The GEM is subsequently used to analyze the processes that affect methods feasibility and evidence quality.

**WHAT IS SUMMATIVE EVALUATION**

Summative assessment term has roots in the field of education which distinguishes it from forma-tive assessment 2. The former assesses students objectively at the end of a study period using standardized exams while the later have a more learning nature about students' progress in meeting the standards. In visualization and visual analytics literature, summative evaluation has been re-ferred to the type of studies that measure the quality of developed solutions. This is contrasted with formative assessment which seeks to learn pros and cons in solution's design.

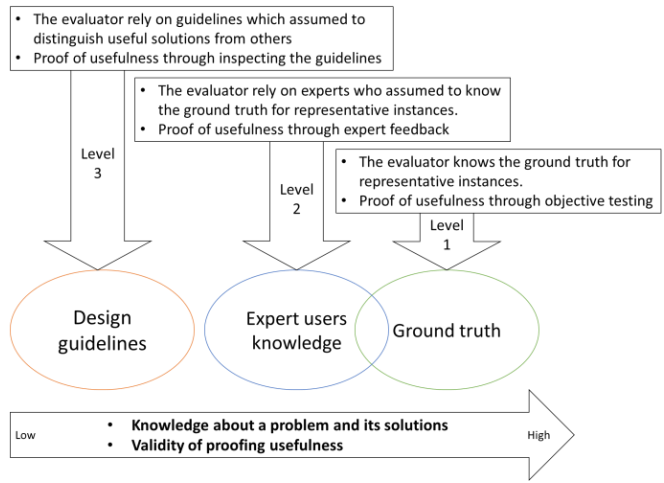

Figure 1. The proposed levels of summative evidence from evaluators prospective.

From the discussion in 2, we derive the definition of summative evaluation as a process which generates evidence (information) about the degree of accomplishment of given objectives (stand-ards) for an assessed object (a solution) at a point in time. Proves can be obtained by measuring the degree of accomplishment of given objectives through objective testing, by collecting qualitative data from interacting with experts or by inspecting guidelines. These three approaches of summa-tive evaluation are ranked according to their validity of proving accomplishment of objectives. Re-searchers apply one of these three approaches according to their knowledge about ground truth as we explain throughout the paper. This ground truth, or gold standard, might be known to evalua-tors in some problems which allow them to conduct formal testing. When the evaluators have lim-ited knowledge about ground truth, they can rely on expert opinion to prove the usefulness of the solution. The experience of experts with a given instances of the targeted problem permit them to authorize useful solutions when ground truth is unknown. Last level of summative assessment relies on inspecting known guidelines or heuristics that supposed to distinguish useful solutions from others. This level has the least validity of proofing usefulness but the highest feasibility as it does not require neither knowledge about the ground truth nor partial knowledge gained through qualitative inquiry with domain experts. Figure 1 Summarize this discussion.

We notice that there are many aliases to the word usefulness in reported evaluation studies. Some researchers use the term "performance" to refer to quantitative assessment and the term "effective-ness" when proving the value of a solution using qualitative methods. Nevertheless, these reported studies have the primary intention to prove the usefulness of a solution and we treat them all as summative evaluation. We also notice many mixing between formative and summative intentions. Many studies apply the second approach of summative evaluation (i.e. expert feedback) to proof the effectiveness of a developed solutions and complement this with formative, learning, processes that inform evaluators of how to improve the design in the future. We think that distinguishing these two types is important to understand the summative part of the evaluation which is used to proof usefulness at reporting time.

**A TAXONOMY OF EVALUATION METHODS**

It is challenging to discuss the process of all evaluation practices that have been conducted in the visual analytics literature. To tackle this challenge, we categorize evaluation methods into 8 main categories (see the leaves of Figure 2). The categories are derived from existing survey papers 3-4 and from reviewing 49 published papers in VAST 2017 that report over 100 evaluation studies. This section presents a taxonomy for evaluation methods with examples for each category from recently conducted studies. The existing taxonomies in the literature 3-4 used to categorize evalua-tion practices according to common evaluation scenarios. Our taxonomy differs from these existing works as we investigate the processes of the evaluation methods themselves, not the context of where they usually applied.

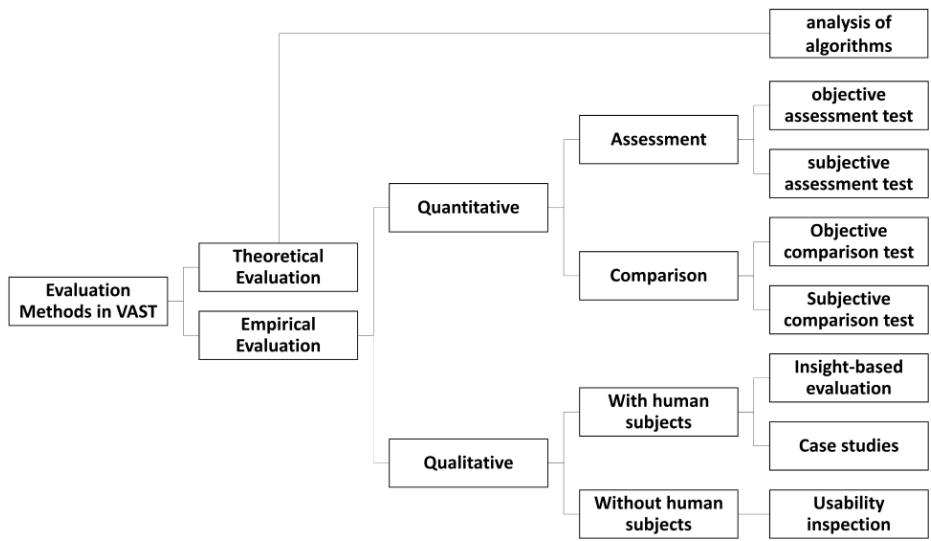

Figure 2. A taxonomy of evaluation methods used in visual analytics literature.

Evaluation methods can either be theoretical or empirical. Theoretical evaluation methods use deductive (rational) reasoning that is developed by relying on certain premises that are logically true. They usually applied at component level which target problems that are measurable and un-ambiguous. Complexity analysis of algorithms is a theoretical evaluation method that can be used to evaluate the efficiency of an algorithm by inspecting its code and measuring the time re-quired to execute its instructions. The analysis reported in 5 is an example of a study that uses this method to measure the efficiency of LOFRCD algorithm which is developed to interactively detect rare categories (groups of outliers similar in their anomalous structure). The findings of such stud-ies hold for every empirical case in general.

Empirical evaluation, on the other hand, follows inductive reasoning by collecting practical evi-dence of usefulness which may hold for a subset of empirical cases. These methods can further be categorized, according to the type of data they collect, into quantitative and qualitative methods. Quantitative methods rely on measurable variables to interpret the evaluated criteria. They collect data in the form of quantities and analyze it using mathematical and statistical procedures to prove their findings. The evidence generated by these methods has a narrow scope but high precision. Thus, these methods are preferable for problems that are well-abstracted to a set of metrics. Quali-tative methods, on the other hand, have less restrictions on the type of data to collect from a study. They evaluate usefulness for problems that are ambiguous or less abstract using data that is less precise but more descriptive such as narratives, voice recordings, interaction logs, screen capturing, etc. That evaluation data can be generated by researchers, as a result of observation, or with the assistance of subjects such as in interviews and self-reporting techniques.

The high precision of quantitative methods gives them the ability to answer two main tasks: as-sessment and comparison. Assessment methods focus on estimating a particular quality in the developed solution. In the objective assessment test, evaluators identify an objective metric or a set of objective metrics that assess the performance of a solution in tackling a problem. This as-sessment is usually employed to measure the performance of the automated algorithmic compo-nents in visual analytics solutions such as outliers-detection algorithms 6. Objective assessment may also test a visual analytics solution holistically. However, a special consideration must be tak-en when estimating the holistic performance of human in the loop solutions to cope with the varia-bility in human performance. An example of a study that applies holistic objective assessment is conducted in 7 to assess a VA solution developed to improve the design and training process of convolutional neural networks. Some assessment methods can be used to estimate subjective met-rics, such as user satisfaction, using questionnaires or structured interviews. An example of such studies is conducted by the authors of 5 who gather subjective opinion using 5-points Likert scale responses about the usability of their interactive rare categories detection solution.

The second type of quantitative methods answers comparison tasks. These methods abstract competing situations into treatments that are defined by measurable variables referred to as inde-pendent variables. The goal of these methods is to uncover if treatments are similar or different with respect to a particular measurable quality referred to as the dependent variable. The dependent variable can be an objective metric such as accuracy. An example is the experiments reported in 8 which compare the accuracy of visual interactive labeling against active learning techniques. The method of labeling is one of the independent variables used in this study to define treatments that were compared with respect to the accuracy of classification, i.e. the dependent variable. In other studies, subjective metric may be used as a dependent variable to compare treatments from sub-jects' perspectives. For instance, the subjective comparison reported in 9 compares Neo4j with a solution called VIGOR which is developed to improve the interpretation of the visual results of graph queries. To ensure that the findings of quantitative comparison methods are valid, treatments must only differ in the levels of the independent variables intended to be studied. It is thus desired to rule out from treatments any factors (confounders) that may become the reason of observing differences in the dependent variable of treatments. Confounders may exist in treatments because of differences in users, tests setup, tasks complexity, etc. In experimental studies, a systematic con-trolled design allows researchers to reduce the effect of confounders. Two common designs are used for this purpose: within-subject (repeated measures) and between-subject (independent measures) designs. In within-subject design, same users are tested at different periods with differ-ent treatments. Between-subject design, on the other hand, randomly assigns users to the treatments to balance human-related factors in the treatments.

Unlike quantitative methods which only are able to generate findings using measurable data that abstract a problem, qualitative methods can generate findings using a variety of data types. These methods define a systematic way of collecting, generating, and analyzing these qualitative data to minimize evaluator's subjectivity that may invalidate the findings generated by these methods. The primary intentions of using these methods are formative or exploratory, but they are extensively used for summative purposes as we explained. Qualitative methods can be categorized into two types. the first type utilizes human subjects to generate findings while the second type completely relies on the evaluator to generate findings without the need of any human subject.

The evaluator in qualitative methods that involve subjects may take active role in generating qualita-tive data such as in case studies. In these methods, a dialog between the evaluator and subjects is established to enable the evaluator from gathering and understanding the information provided by the subjects who are usually domain experts. The evaluator can utilize many data gathering tech-niques such as semi-structured interviews, think-aloud protocol, interaction logging, observation, etc. The evaluator analyzes collected data and ensures it accurately describes the studied cases and answers the evaluation questions. The case study reported in 10 shows the usefulness of Voila, a system that enables domain expert from detecting and analyzing anomalies in spatiotemporal data. Such studies show how evaluators can rely on domain expert subjects to validate the usefulness of a solution. Case studies can come in different variants. In traditional case studies, a domain expert uses a solution and report any limitation or usefulness in it. The domain expert may be assisted by the evaluator when the goal of evaluation is to study expert reasoning rather than the solution's usability. This appear in cooperative evaluation methods such as Pair analytics 11 and MILC 12.

A special type of evaluation methods that utilizes subjects to collect qualitative data is the insight-based evaluation method 13. This method asks subjects, who use a solution to tackle analysis problem, to self-report any insights they reach during the analysis by applying techniques such as diary or think-aloud protocol. The insight-count can then be used as an objective metric that can be used for assessment or comparison goals. This quantitative analysis of qualitative data makes this method unique. The unobtrusiveness of evaluator during data collection is another factor that dis-tinguishes this method from other qualitative methods.

Usability inspection methods validate the effectiveness of a solution without testing or recruiting subjects. Unlike testing methods previously mentioned, inspection does not require an actual im-plementation of evaluated system and can be applied during the design phase. In these methods, evaluators inspect a solution to check the satisfaction of predefined requirements. We categorize these methods under qualitative categories because evaluators derive usefulness evidence from inspecting a design, a clear qualitative form of data. Heuristics, or design guidelines, are a special type of usability requirements that have been developed from community-level experience and are broadly accepted to be useful. Inspecting these heuristics is a well-known evaluation method re-ferred to as heuristic evaluation 14. Another inspection method is the cognitive walkthrough 15 which assumes that the evaluator knows usability principles and thus can directly inspect a design without explicitly predefine a set of heuristics.

## THE GENERIC EVALUATION MODEL (GEM)

We propose a process-based generic evaluation model (see Figure 3). The model is developed by inspecting the 8 categories of evaluation methods in the taxonomy we have discussed. The goal of developing the model is to highlight factors that affect the feasibility of evaluation methods and the validity of the evidence they provide. This allows us to explain why researchers select one method over another to prove usefulness. The model is generic as it broadly applied to any type of solu-tions, including manual, automated, and human in the loop solutions. In this section, we explain the abstraction of evaluation process and the functionality of GEM components in terms of the process they execute. We illustrate how to utilize this architecture in the next section to perform validity and feasibility analysis of the evaluation methods.

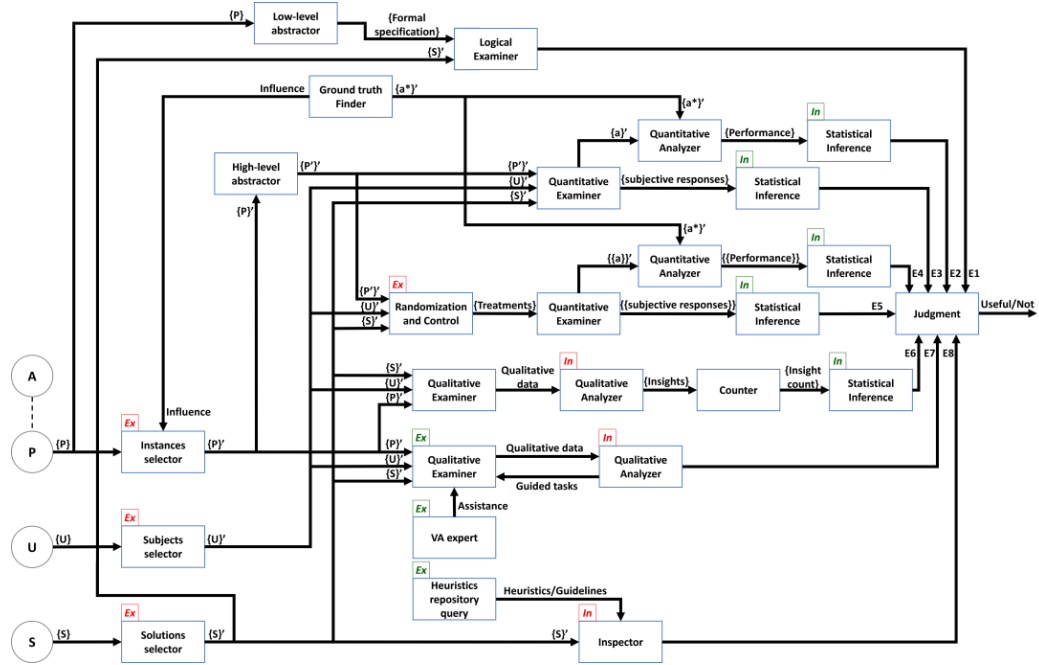

Figure 3. The Generic Evaluation Model Shows a general process of evaluating usefulness of a solution. Different evaluation methods take different paths in GEM and end with evidence that are used to judge the usefulness. Green and red labels highlight processes that have the potentials to affect the validity of the generated evidence positively or negatively.

### Characterizing GEM's End-to-End Evaluation Process

In this section we characterize the E2E evaluation process. The process has three inputs: the set of problem instances {P}, the set of solutions {S}, and the set of targeted users {U}. These input sets form the spaces P, S, and U respectively, and are uniquely defined for every problem. Each prob-lem instance is associated with an answer a* which is desired when tackling that problem instance. The set of correct answers for all problem instances is denoted by {a*} which is a set defined in the answer space A. The output of the evaluation process is a decision about the usefulness of the evaluated solution which is made according to the generated evidence of one of the evaluation methods.

To evaluate the usefulness of a solution, one needs to evaluate how accurately it can solve the prob-lem it was developed to tackle. Problems are obstacles that create challenges to reach a desired goal or objective. Munzner 16 defines four levels of abstraction of these problems: high-level domain problems, lower-level domain problems, high-level abstract problems, and low-level abstract prob-lems. Visual analytics solutions should be evaluated at each level of abstraction as suggested by Munzner's nested model. During the development of the GEM, we carefully consider this argu-ment in the abstraction of the problem concept. We treat each level of abstraction as a completely independent problem with a completely different set of solutions. For example, VA solution may target a sorting problem at a low abstraction level and when evaluating the solution at this level, the GEM represents sorting algorithms as solutions for that abstract problem. Moving the problem to other abstraction level, e.g. user interpretation of sorting algorithm results, explicitly means that the evaluated solution S in this new evaluation problem changes as well, e.g. to human-based solu-tions. This idea is clarified well by the concept of blocks proposed in 17. To meet our objective of understanding the process of evaluation methods themselves, we design the GEM by abstracting the concepts of problems and solutions and then evaluate different problems independently.

The set of problem instances {P} contains all possible instances of a problem. Problem instances are cases of that problem and are defined by different context parameters. We may have limited knowledge about the context parameters which may impact our abstraction and problem-solving processes. One of the main roles of qualitative inquiry is to help us increase this knowledge and understand the problem better, such as in pre-design qualitative studies 18. In visual analytics, prob-lem instances are usually distinguished by the datasets used in the analysis. An example of a prob-lem at a domain level is selecting the optimal response to control an epidemic spread. This problem can have many instances that can be characterized by different geospatial locations, temporal peri-ods, demographics, etc. The answer space for a problem contains all possible answers that can be chosen when attempting to solve a problem instance. In our previous example, A contains all pos-sible sequence of actions that can be taken by the authorities to control the epidemic, including quarantining, vaccination, etc. Solving a problem is a search for a mapping function F: P → A that maps problem instances to their answers which work in general, i.e. correctly maps most of the instances.

Many solutions may be developed to tackle a problem. Solutions characterize the mapping function F according to the assumptions, features and the vision of their developers. Human in the loop solutions such as visual analytics solutions are special types of these mapping functions which take a set of users {U} from the user space U as an input. Users may generate the decisions for problem instances or can control the behavior of the automated algorithm which performs the mapping. An example of visual analytics solution that tackles the epidemic problem previously mentioned is presented in 18. In that tool, users are provided with a set of actions that can be taken at different phases of the epidemic to control its spread. The tool then predicts the effect of these actions based on a simulation model of epidemic spread which assists users in evaluating different plans of ac-tions they hypothesize according to their domain knowledge. Note that multiple solutions may exist and the set {S} contains all developed solutions that can be used to tackle the problem P.

The output of the GEM is a decision about the usefulness of the evaluated solution. This decision requires evidence to prove that the evaluated solution is indeed useful. This evidence is generated by different evaluation methods that can be represented by different paths in the GEM. The afore-mentioned eight categories of evaluation methods provide 8 types of usefulness evidence, which is symbolized by E1, E2, …, E8. In the following sections, we illustrate these types of evidence and how to reach them.

**Component-level processes in the GEM**

We start to examine the eight categories of evaluation methods to unfold the processes that distin-guish them from one another. This allows us to explain differences between the categories and helps us in our feasibility and validity diagnoses.

In theoretical evaluation, the evaluator applies two processes in sequence: low-level abstractor and logical examiner, which generates the first type of evaluation evidence E1. The low-level ab-stractor process abstracts the problem to a set of formal specifications. Such abstraction requires sufficient knowledge about the problem to represent it using a formal language with a defined syntax and semantics. The formal verification is then conducted in the logical examiner process to confirm the correctness of the solution. The logical examiner basically checks the logic of the solu-tion and determines if this logic matches the logic of the formal specification. Details about this type of evaluation can be found in 20. In cases where computational complexity is the intention of the evaluation instead of correctness, no formal specification is needed. The logical examiner in these cases examines the logic of the solution to determine the resource complexity of the solution, for instance, in terms of worst case scenario. Resources can be execution time or required memory space and they are defined as a function of solution's input size.

Unlike theoretical evaluation, empirical evaluation relies on sampling to empirically evaluate a solution. Selector processes are responsible of electing cases, competitors, and participants in-volved in an evaluation study. Instances selector is the process employed in a study to define cases to be included in the study. It is a sampling process which selects sample cases from the cases that can be observed in real world. The output of this process is a set of problem instances to be studied {P}'. Solution selector is the process of selecting a solution or a set of solutions {S}' from all of the available solutions developed to solve a problem. The solution selector selects a single solution, i.e. the evaluated one, when the intention is assessment or inspection. The selector may select multiple competing solutions in comparison evaluation methods. Subjects selector process is responsible for selecting the set of users {U}' involved in the evaluation study from the set of all targeted users. Users are required to evaluate human in the loop solutions.

**Quantitative evaluation tracks**

In quantitative evaluation methods, the selected problem instances are abstracted to a set of well-defined tasks with measurable outcomes. This abstraction is done in the high-level abstractor process. The tasks try to characterize original problem and thus measure objective qualities. The tasks may quantitatively examine subjective qualities as well which assumed to correlate with the objective usefulness. The difference between low-level and high-level abstractor appears in the required amount of knowledge to perform the abstraction. The former requires knowledge of both the problem and the behavior of the correct answer which is important to develop the formal speci-fication. High-level abstractor, on the other hand, requires less knowledge that only allows accurate quantitative description of the problem and the answer spaces. The set of abstracted sampled prob-lem instances, denoted by {P'}', and the corresponding correct answers, denoted by {a*}', is then used in quantitative methods to generate empirical evidence of usefulness for examined solutions.

Once the problem is reduced to a set of measurable objectives, the researcher has the choice to assess a solution alone against a pre-defined baseline or against competitor solutions. In both choices, solutions are tested in the quantitative examiner process. In this process, solutions are tested with the set of tasks and provide either objective or subjective responses. To be able to apply objective evaluation methods, the evaluator needs to know the correct answer for the selected prob-lem instances. These correct answers are learned or prepared by the evaluator for some problem instances upfront in the ground truth finder process. The knowledge of the evaluator about the ground truth of some problem instances may influence the sampling of problem instances as in supervised learning experiments. In these experiments, evaluators utilize labeled data (i.e. instances with known answers) to evaluate prediction models using methods such as cross-validation or hold-out test set to predict the general performance 21. In other cases, evaluators may mimic prob-lem instances and generate synthetic instances with known ground truth as suggested in 22. The ground truth is used in quantitative analyzer process which generates performance scores using similarity measures between objective responses and the ground truth.

In studies with comparison intension, randomization and control process is responsible for con-structing the treatments involved in the study according to study design. In experimental studies, this process is responsible of controlling or ruling out the effect of any confounder by randomly assigning subjects to groups. This process distinguishes experimental studies from quasi-experimental studies which has less control of confounders.

Evaluators may apply inferential statistical test to the generated scores to support the empirical evaluation evidence with statistical power. Methods such as Analysis of variance (ANOVA) is an example of the inferential statistical test which is commonly applied in comparison studies to statis-tically prove that the difference between the means of treatments scores is not observed because of an error in sampling. In assessment studies, statistical inference process may generate the confi-dence interval for the measured performance. Subjective treatment responses may undergo similar statistical inference procedures in subjective comparison and assessment methods.

Four types of empirical evidence can be generated from quantitative evaluation tracks. E2 and E3 are the solution performance and the subjective responses scores that assessment methods gener-ates and can be used as empirical evidence of usefulness. Similarly, E4 and E5 are the difference in the sample distribution of treatments' objective and subjective scores that are generated by compar-ison methods and can be used as empirical evidence of superiority.

**Qualitative evaluation tracks**

Qualitative methods take a different direction to validate the usefulness of developed solutions. Instead of abstracting the problem instances to particular objectives, qualitative methods study problem instances in their original form which may contain vague, less-structured tasks. To extract evidence of usefulness from such studies, evaluators pay extra attention to any type of data that can be captured during the examination which is emulated in the qualitative examiner process. Data in these methods can educate the evaluator about different factors in the solutions as well as the value of them from user's perspective. Many methods can be implemented in this process to extract and collect qualitative data including: observation, semi-structured interviews, subjects' feedback, Think-aloud protocol, video and audio recordings, interaction logs and screen capturing 23.

The resulted qualitative data from the examiner process is analyzed in the qualitative analyzer with the aim to increase the knowledge of the evaluator about the quality and the effectiveness of the solution.

the solo role of this process in insight-based evaluation is to identify the insights reported by the users through think-aloud or diary methods. These insights are then counted for each user, and the evaluator may use this count as a quantitative measure of usefulness. This meas-ure can be used as an assessment metric or can be used to compare multiple solutions. The insight-counts, as scores, can then undergo inferential statistical tests to prove generalizability as in quanti-tative methods. The average/total insight-count for a solution or the superiority of one solution over another in terms of average/total insight-count is the empirical evidence of usefulness, E6, that is generated by insight-based evaluation method.

In case studies, the qualitative analyzer may work in parallel with the examiner process to guide the data collection methods. A visual analytics expert may provide assistance to the users on how to use the solution to let them focus on the studied cases that may be impacted due to solution's steep learnability curve. The role of VA expert is assigned to the evaluator in most studies. The process of data collection and analysis continues, according to qualitative research logic, until no further information can be extracted about the studied cases. The collected information from these studies can then work as an evidence of how useful a solution is in users' eyes E7.

The evaluator may rely on inspection methods to evaluate the usefulness of a solution. In these methods, the inspector process generates evidence of usefulness E8 according to the evaluator's opinion who tries to walk in users' shoes to predict solutions usability. The evaluator may utilize a pre-developed heuristics and design guidelines which can be queried from the literature. Usability correlate with usefulness as reported in 24.

## ANALYSIS OF VALIDITY AND FEASIBILITY

The GEM model allows us to highlight processes that limit the applicability of evaluation methods and processes that affect the validity of the evidence they provide. This can inform practitioners about the potential risks that one may face when using one method over another to assess the use-fulness of a solution. Two types of validity are discussed with explanation of why certain process-es affect them. After that, we pinpoint some processes that is not feasible for some problems. both analyses then can lead to a solid reasoning of why researchers select a certain evaluation method to evaluate solutions of a certain type of problems.

### Internal Validity and External Validity

Many types of evidence validity can be used to assess the quality of evaluation studies from differ-ent angles 25. We select two types of validity and explain how to generalize their interpretation to be able to analyze them in the GEM. These validity types are Internal validity and External validity. Our interpretation of these two types is derived from the original work that introduced them by Campbell and Stanley 26, and the work proposed by Lincoln and Guba 27 who attempt to apply them in qualitative research context.

Internal validity is commonly known as the validity of causation findings in experimental studies. It validates that there is a cause-and-effect relationship between the independent variable and the de-pendent variable. External validity, on the other hand, is used to check for generalizability of find-ings to other actors and situations. Regardless of their traditional definitions that limit their applica-tion to experimental studies, their concepts may be viewed with a larger scope. For instance, Lin-coln and Guba apply the logic of internal and external validity in qualitative research by proposing the terms credibility of subjects and transferability of results to other cases. following this general logic allows us to redefine the two validity types. To explain the definitions, we introduce the con-cepts of study population space and samples representativeness space.

A study population space is the space that describes every possible evaluation case that can be characterizes by the spaces P, S, and U. When an empirical study is conducted, a sample from the cases is selected which defines a sample representativeness space that either represents the whole population space or a subspace of it (according to the study hypothesis). The results of a study may be applicable to all cases in the representativeness space or to a part of it. We view external validity (Ex) as the extent of how representative a sample space is to the population space, a measure of generalizability. On the other hand, we view internal validity (In) as the extent of how well study findings apply to the sample space, a measure of correctness. A visual illustration of this discus-sion is presented in figure 4.

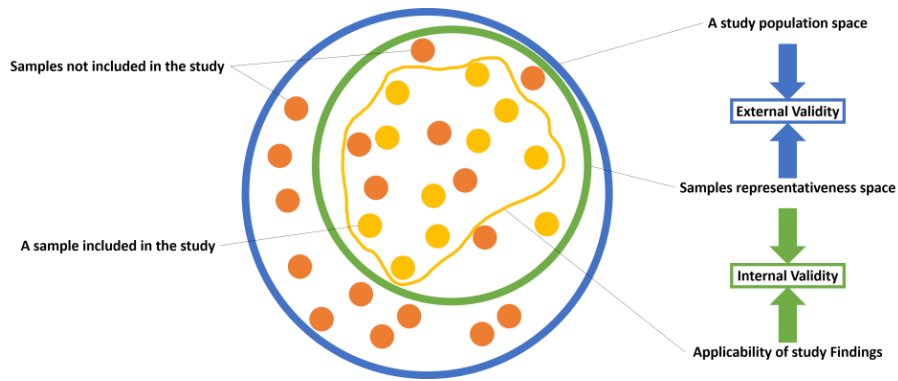

Figure 4. An illustration of our view of internal and external validity. Internal validity is concerned with the applicability of study findings to the cases that are represented by the study settings. External validity is concerned with the applicability of study findings to the population cases in general.

**Analysis of Evaluation Methods**

Theoretical methods do not build evidence of usefulness from samples but from considering the whole population space. thus, there is no concern for external validity in these methods as they provide completely generalizable evidence by nature. Any theoretical framework is also proved to generate valid evidence if applied correctly. For example, measuring the time complexity of an algorithm can generate evidence of the upper-bound of the time needed for an algorithm to be exe-cuted. This upper-bound is provable to be a correct characteristic of the worst-case scenario of execution time. However, theoretical methods are the least feasible methods. Low-level abstrac-tion of a problem is very costly as it requires extensive analysis of the problem. The complexity of developing formal verification framework for interactive systems reduces the applicability of this method. In fact, we have not encountered any study that apply formal verification method in VA literature because of this challenge. For efficiency assessment, theoretical methods are only capable of evaluating automatic algorithms because of the infeasibility of rationally reason about human's ambiguous analysis and problem-solving processes.

The existence of a human in the loop forces evaluators to move from rational to empirical reasoning to prove usefulness. Empirical evaluation methods rely on samples to generate usefulness evidence, which may introduce a risk to the external validity of the findings. This risk may occur in empirical methods due to bias in sampling from the population space, i.e. bias in the selector processes.

Quantitative methods focus on proving usefulness with evidence that is internally valid even if this cost a reduction in external validity. One example appears in quantitative comparison methods which traverse the randomization and control process. This process may reduce the external validity of comparison methods by creating unrealistic cases as a result of controlling factors not intended to be studied. The control aims to eliminate the effect of confounders to maintain internal validity at the cost of external validity.

The precision of quantitative evidence enable evaluator to prove their correctness through statisti-cal inference tests. These tests can improve the internal validity by examining the statistical signif-icance of findings to generalize them to cases in the study space. However, quantitative methods require high-level abstraction of the problem which may not be feasible. This abstraction requires developing a set of tasks that accurately represents the problem while enables collecting measurable evidence of usefulness. Objective evidence is also less feasible than subjective evidence because they require knowledge about the correct solution for evaluated cases, i.e. the ground truth.

Qualitative formative assessment is used when it is not practical to perform the high-level abstrac-tion or ground truth finder processes. These methods are more feasible but usually focus on specific cases, i.e. may have small study space. The experience of subjects, however, make up for this limitation and may increase the applicability of study findings for cases outside the study space. Qualitative evaluation methods may not benefit from statistical inference as quantitative methods do, but they have the advantage of utilizing the knowledge of subjects. This benefit may appear in the qualitative examiner process which can apply less structured tasks to gather data that have larger scope than the direct tested cases. This advantage is useful in formative, descriptive and exploratory intentions, which aim at generating hypotheses instead of testing them 28. Howev-er, for summative intention, this flexibility comes with costs in internal validity. The semi-structured tasks presented to subjects in the qualitative examiner may be interpreted differently with different subjects which subsequently affects users' behavior and opinion. Moreover, the qualitative analyzer process may reduce the internal validity of the concepts it generates as a result of evaluators' subjectivity in understanding qualitative data. Both issues can be overcome by experience of conducting this type of qualitative studies and following guidelines that reduce sub-jectivity such as those proposed in the grounded theory 30. Methods such as Triangulation and Member-checking can be utilized to ensure that the qualitative findings are internally valid, i.e. cred-itable. However, no formal mathematical prove can be done with qualitative data.

A special case of qualitative methods is insight-based method. This method assesses the usefulness of solutions by defining summative metrics such as insight-count. The internal validity of these metrics can be supported by statistical test. However, the drawback in insight-based method ap-pears in the qualitative data collection techniques, as previously mentioned.

The most feasible evaluation methods are usability inspection methods which do not even require testing processes. This high feasibility makes these methods suitable for evaluation in early devel-oping phases when modification and re-evaluation have high frequency. However, these methods prove usefulness marginally and with many validity concerns. Validity of inspection findings total-ly relies on the evaluators and thus characterizes their skills. Internal validity describes evaluators correctness of capturing usability flaws in the evaluated solution. External validity, on the other hand, describes their skills in detecting usability issues in general which can be linked to experience and can be supported by following guidelines and pre-defined heuristics.

## USE CASE

In this section, we walk through three evaluation studies reported in 31 and map them to the GEM as illustrative examples. That paper proposes ConceptVector, a visual analytics system that guide users in building a lexicon for custom concepts and use them in corpus analysis. The authors de-fine a concept as "a set of semantically related keywords characterizing a specific object, phenome-non or theme".in the following discussion, any bold text is a component in the GEM (see figure 3).

The first reported evaluation task seeks to assess the usefulness of ConceptVector's capability of building a lexicon for a concept. The study compares ConceptVector objectively against two coun-terparts. The problem space P for the first evaluation problem contain all possible words in English language, so does the answer space A. the solutions for this problem maps each word from P to a set of words in A that supposed to share same semantics. This problem cannot be evaluated theo-retically because it is not feasible to derive formal specification of interactive systems nor defining premises for the time that human analysis takes. Thus, authors were forced to evaluate the solution empirically. From the problem space, the authors selected three problem instances that represent three English words {P}' = {"family", "body" and "money"}. Three solutions are selected to be part of this evaluation {S}' = {ConceptVector, WordNet 32, Thesaurus.com}. Fifteen computer science graduate students with high computer skills are sampled to represent targeted users {U}'. the ground truth process for this study utilizes the Linguistic Inquiry and Word Count 33 to gen-erate the sets of words {a*} that are semantically related to each of the three problem instances. It is feasible for this problem to be abstracted to a set of tasks that can be assessed with measurable quality. This allows authors to conduct quantitative evaluation. Because the authors are interested in comparison evaluation, they conduct randomization process which randomly assigned users to the three solutions for each problem instance. The Users then build a lexicon for the given concepts in the quantitative examiner and the answers compared with the {a*} in the quantitative ana-lyzer. The analyzer uses precession, recall and number of words as an objective metrics. Statistical significance tests which utilizes pairwise Tukey HSD method are applied to the scores and the results impower the evidence E4 that is used to prove usefulness.

The second evaluation task that was tackled by the authors was to prove of the usefulness of their automatic ranking algorithm that rank keywords according to their relevance to a developed con-cept. A problem instance in this problem is characterized by a concept and a set of keywords. The answer for a problem instances is a ranked list of the keywords according to their relevance to the given concept. Using the ground truth finder process, the authors found a list of 10200 key-words that is manually ranked according to their relevance to the concept "Happiness". This find-ing influences the selection of the problem instance that is characterized by the concept "Happi-ness" and a set of keywords taken from the manually ranked keywords. The selected solutions {S}' to be compared in this evaluation study was 6 automatic methods of ranking that differ in word embedding technique and ranking algorithm. Comparing automatic solutions does not require any users and thus {U}' = Φ. Similar to the first problem, this problem can be evaluated with quantitative methods. The absence of users, who are a source of variability, renders the randomi-zation and control process idle. The algorithms rank the

keywords in the quantitative examiner process and report the ranking to the quantitative analyzer that measure the Spearman's correla-tion between the algorithms ranking and the ground truth, i.e. the manual ranking. No statistical significance test is needed in such studies that does not have variability in the outcome of the quantitative examiner process. The evidence E4 is again used to prove the usefulness of authors KDE-based ranking solution with respect to competing methods.

Finally, the authors also report a third evaluation study which evaluate the user interface of Con-ceptVector in a formative manner. The primary problem that is targeted by ConceptVector is to develop concepts and then use them to analyze a corpus. Thus, the problem instances tackled by ConceptVector are collections of text documents. In the third study, the authors select New York times comments as the instance for this experiment. Analysis problems have the objective of deriv-ing insights from a given data and thus, the answer for the selected instance is the set of all insights that can be derived from the New York comments. Selected solutions that assist tackling this prob-lem are {S}'={ConceptVector, Empath 34}. For this complex problem, it is not feasible to evaluate solutions theoretically nor objectively because it is not feasible to abstract the problem into tasks that accurately represent it and can be assessed quantitatively. This force the selection of qualita-tive method to prove usefulness. Because one of the authors' tasks was to gather users' feedback about the system design, expert review was the evaluation method of choice. This method can be considered a special case of case studies but with less involvement from qualitative analyzer (i.e. evaluators). During the qualitative examiner process the users keep sending information to the evaluators' qualitative analyzer process about the features that are useful in ConceptVector as well as other feedback. For example, the users reported the usefulness of showing recommended words using scatterplot over traditional list. Evaluators collect the feedback, analyze it and then conclude with information that represent users' opinion about the system. This opinion is then used as a qualitative evidence of usefulness E7.

From these three examples, we can observe the flexibility of the GEM architecture which is able to capture the main processes in different evaluation studies. The provided cases also show how do we reason about the selection of an evaluation method for a given evaluation task.

**CONCLUSION**

This paper provides an entry point to researchers who are interested in solving the challenge of evaluating visual analytics solutions. We take an overview of summative evaluation methods that have been applied in visual analytics literature. By abstracting the processes of evaluation, we were able to inform readers about the main components that impact the feasibility of evaluation methods and the quality of the evidence they provide. Each highlighted validity or feasibility concerns can be considered a research challenge and can be tackled in the future to improve current evaluation prac-tices of visual analytics solutions.